# Deep Learning-Based Channel Squeeze U-Structure for Lung Nodule Detection and Segmentation

1st Mingxiu Sui
University of Iowa
Iowa City, USA

2nd Jiacheng Hu
Tulane University
New Orleans, USA

3rd Tong Zhou
Rice University
Houston, USA

4th Zibo Liu
Johns Hopkins University
Baltimore, USA

5th Likang Wen
Johns Hopkins University
Baltimore, USA

6th Junliang Du*
Shanghai Jiao Tong University
Shanghai, China

*Abstract—This paper introduces a novel deep-learning method for the automatic detection and segmentation of lung nodules, aimed at advancing the accuracy of early-stage lung cancer diagnosis. The proposed approach leverages a unique "Channel Squeeze U-Structure" that optimizes feature extraction and information integration across multiple semantic levels of the network. This architecture includes three key modules: shallow information processing, channel residual structure, and channel squeeze integration. These modules enhance the model's ability to detect and segment small, imperceptible, or ground-glass nodules, which are critical for early diagnosis. The method demonstrates superior performance in terms of sensitivity, Dice similarity coefficient, precision, and mean Intersection over Union (IoU). Extensive experiments were conducted on the Lung Image Database Consortium (LIDC) dataset using five-fold cross-validation, showing excellent stability and robustness. The results indicate that this approach holds significant potential for improving computer-aided diagnosis systems, providing reliable support for radiologists in clinical practice and aiding in the early detection of lung cancer, especially in resource-limited settings*

*Keywords-Deep Learning; Lung Nodule Detection; Early Diagnosis; IoU*

## I. INTRODUCTION

Lung cancer is a lethal malignancy worldwide. The 5-year survival rate for stage I lung cancer smaller than 1 cm can be as high as 92%, while the survival rate for advanced lung cancer remains extremely low, with patients theoretically surviving only a few years. The majority of lung cancer patients are diagnosed at an intermediate or advanced stage, leading to poor prognosis and low survival rates. Early diagnosis and detection are critical challenges in this field. Moreover, the latest cancer statistics report indicates that lung cancer is expected to account for 21% of all cancer deaths in the United States in 2023, with 127,070 people dying from lung cancer [1]. Even more concerning, over half of the world's lung cancer cases occur in developing or underdeveloped countries [2], where medical resources are more limited than in developed nations.

Currently, low-dose CT is the most important screening method for lung cancer and is widely used for diagnosis, treatment planning, efficacy evaluation, and prognosis observation, significantly reducing lung cancer mortality [3].

However, the symptoms of early lung cancer are not obvious, and pulmonary nodules are the main manifestations of lung cancer, which are generally opaque circular anomalies with a diameter between 3mm and 30mm [4]. The US National Cancer Institute developed the Lung Image Database Consortium and Image Resource Initiative (LIDC) dataset for research into screening of lung nodules [5]. Correct localization and segmentation of pulmonary nodules assist doctors to preserve the normal functional tissues and organs during treatment. Detection and segmentation of pulmonary nodules require doctors to locate pulmonary nodules in hundreds of slices and manually outline the lesion area. Secondly, due to the small lesion area of pulmonary nodules compared with the area of the CT section, high-intensity, and long-term CT imaging is highly likely to cause missed detection of small nodules, imperceptible nodules, or ground glass nodules, which may eventually lead to misdiagnosis.

To solve the above problems, the computer-aided diagnosis system has been produced and developed rapidly, especially the diagnosis method based on deep learning technology[6-8]. The method is based on the big corresponding paradigm to realize the segmentation of lesion detection. In order to achieve accurate segmentation of pulmonary nodules, this paper proposes a pulmonary nodules segmentation method based on channel residual U structure. In this method, three modules of shallow information processing U structure, channel residual structure, and channel extrusion U structure are proposed. Firstly, the shallow information processing U structure is used to enhance the interaction and extraction of feature information between different feature layers. Finally, the channel extrusion U structure is used to integrate the feature information of different semantic levels. Through the verification of the network structure and the final overall experiment results, it is shown that our approach holds significant potential for improving computer-aided diagnosis systems.

## II. RELATED WORK

Deep learning has notably transformed medical imaging, significantly advancing lung nodule detection and segmentation through innovations in neural network architectures, data augmentation, and multimodal fusion. Among these developments, the integration of Generative Adversarial Networks (GANs) has been pivotal. For instance, Feng et al. [9] highlighted the role of GANs in generating realistic medical images from limited datasets, a technique that enhances model training and generalization—critical for our work on synthetic nodule creation. In parallel, the application of multimodal data fusion has been recognized for its potential to elevate diagnostic accuracy. Liu et al. [10] discussed how combining diverse data modalities, such as imaging and clinical data, can refine decision-making in disease recognition. Although our current focus is on CT images, exploring multimodal fusion could be an intriguing future direction. Preprocessing techniques also play a crucial role, as evidenced by Xu et al. [11], who emphasized their significance in preparing medical datasets for deep learning applications. Their strategies for enhancing model performance through sophisticated preprocessing align closely with our methods for CT scan preparation.

Moreover, Convolutional Neural Networks (CNNs) continue to be a cornerstone in medical image analysis. Xiao et al. [12] applied CNNs to cancer cytopathology image classification, an approach analogous to ours in lung nodule detection, where both tasks demand precise feature extraction from intricate image data. Further extending the utility of GANs, Zhong et al. [13] and Yang et al. [14] explored their effectiveness in image recognition and diagnosis. Yang et al. demonstrated how conditional GANs (cGANs) could produce more accurate imaging diagnostics, a concept that could enhance our synthetic nodule generation, tailoring it to specific clinical characteristics. Innovations in deep learning architectures also contribute to better data handling and feature representation. Cheng et al. [15] combined ELMo word embeddings with deep learning-based multimodal transformers for image description tasks. This advancement in extracting and representing complex features is adaptable to our lung nodule detection framework. Attention mechanisms in deep learning, such as those employed in the Attention-Unet model by Zhu et al. [16], have shown exceptional promise in improving segmentation accuracy. Our proposed "Channel Squeeze U-Structure" incorporates similar principles to enhance segmentation through more effective feature integration. Understanding intra-tumor heterogeneity is crucial, as explored by Wang et al. [17] in their study on temporal mutations in cancer pathways. Although their focus was on genetic pathways, the principles of capturing variation in medical data are pertinent to analyzing the diverse characteristics of lung nodules, including aspects like ground-glass opacity. Deep learning's prowess extends to survival prediction across different cancer types, as demonstrated by Yan et al. [18]. Their methodology, focused on broad cancer diagnosis, reinforces the potential of neural networks to make accurate predictions, resonating with our paper's objective to harness advanced neural networks for robust healthcare applications. Furthering the discourse, Gao et al. [19] explored the optimization of text classification using graph neural networks, shedding light on techniques that could be adapted to medical imaging. Their approach suggests that optimizing neural network performance could be beneficial for structural analysis in image segmentation, similar to our proposed network structures.

In a novel application of hypergraphs, Yang et al. [20] developed a method to exemplify the utility of complex graph-based models in medical data analysis, offering insights that could be adapted for improving the detection and segmentation of lung nodules. Zheng et al. [21] introduced adaptive friction into deep learning optimizers, presenting a method to refine the learning process. The implementation of such advanced optimization techniques could improve the stability and convergence of the neural networks used in our lung nodule segmentation tasks. Enhancements to convolutional neural networks (CNNs) were also addressed by Wang et al. [22], who integrated higher-order numerical difference methods to boost feature extraction. This refinement is particularly pertinent to our work on enhancing U-structure networks for precise segmentation of intricate lung nodule structures. Optimization strategies were further explored by Ma et al. [23] and Liu et al. [24], who delved into advanced gradient descent techniques in neural network training. These strategies could augment the efficacy of our channel residual and squeeze structures, ensuring robust performance even with complex datasets like lung CT images. In the realm of NLP, Zheng et al. [25] emphasized the value of pre-trained models, a strategy that could be translated to image segmentation to enhance lung nodule detection models, drawing parallels in methodology improvement across different domains of machine learning[26]. Liang et al. [27] presented an unsupervised image registration method using Dense U-Net and channel attention mechanisms. Their findings are directly applicable to our proposed model, which also utilizes channel attention to optimize feature integration, thereby improving the accuracy of detecting small structures in medical imaging. Lastly, Li et al. [28] combined knowledge graph embedding with deep learning to develop a model for predicting adverse drug reactions. Although focused on pharmacology, their approach highlights the potential for integrating external medical knowledge into deep learning models, suggesting possible enhancements for lung nodule detection through enriched decision-making processes.

These studies collectively enrich our understanding and application of deep learning technologies in medical imaging, paving the way for innovative and effective approaches in lung nodule detection and segmentation.

## III. ALGORITHM PRINCIPLE

The Channel Residual U2Net 3D network proposed in this paper based on the UNet network is shown in Figure 1. The network contains two stages of information encoding and decoding, and the network feature map can be expressed as $x$. The dimensions of the input and output feature maps are $C \times H \times W \times D \in 1 \times 64 \times 64 \times 64$. The encoder is composed of En and $i \in [1,4]$. The inside of the encoder is SIPU and the Channel Residual U-structure. The SIPU module is used to balance shallow feature information, and the CRSU module is used for information encoding and transmission. The input resolution of the first four stages is 64×64×64, 32×32×32, 16×16×16, and 8×8×8, respectively. The number of channels in the network also increased from 1 channel to 32, 64, 128, and 256. In the bottleneck layer, the module and CEU module are designed to sense and optimize the feature information transmitted in the first four stages. The input feature map size of the bottleneck layer is 256×4×4×4.

The decoder consists of Dei,i∈[1,4], four parts, which are also filled by SIPU module and CRSU module in turn. The basic block of the decoder is the same as the basic block of the encoder in the same layer. The input feature map of the decoder basic block is composed of the output feature map of the encoder basic block of the same layer and the output feature map of the decoder basic block of the next layer. Take the basic block input of De2 as an example, the input is the concatenated feature map from En2 encoder and De3.

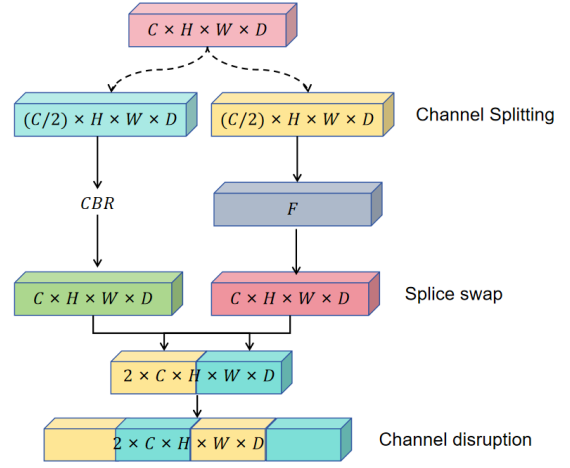

Figure 2 Channel residual structure

Finally, this paper gives the loss functions used, which can further reduce the difference between the true value and the predicted value, so that the model can obtain better segmentation results.

$$L_{CE(P)} = -\sum_{C, y_{p,c}} \log(y'_{p,c})$$

$$L_{Dice} = 1 - \frac{2\sum_{p \in P} y'_p y_p + \varepsilon}{\sum_{p \in P}(y'_p + y_p) + \varepsilon}$$

Where C is the category index, $y_{p,c}$ is the true label of pixel p in category c, and $y'_{p,c}$ is the probability that pixel p is predicted to be of category C. In the second formula, $\varepsilon$ is a small positive number to avoid division by zero.

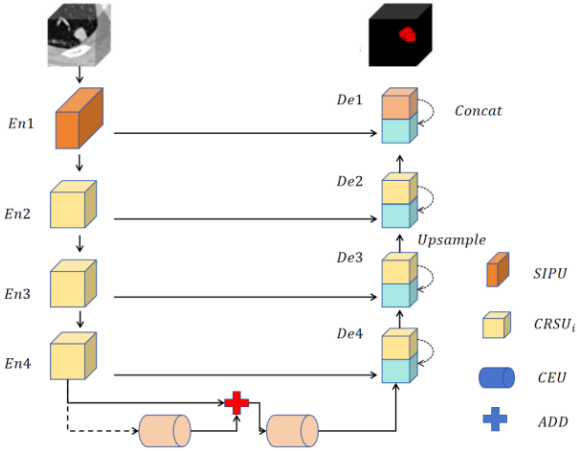

Figure 1 Network architecture diagram

This paper proposes the Channel Residual Structure (CR), analyzing its similarities and differences with the classic residual structure from a structural perspective. Residual connections not only effectively alleviate the vanishing gradient problem but also enable the network to focus on features that need abstraction. The computational process of a residual connection can be represented as:

$$x' = F(x) + x$$

where $F$ is the weighted layer responsible for feature learning and optimization, $x$ on the right side is the original feature map undergoing identity mapping, and $x'$ on the left side is the output feature map, which is the sum of the optimized weighted layer and the original input feature map xx. The channel residual structure proposed in this paper is shown in Figure 2.

IV. EXPERIMENT

A. *Experimental setup*

The experiment is based on the Ubuntu 18.04.3 LTS operating system. The computer's brain is an Intel (R) Gold 6140, it's got 187G of brainpower, and it runs on Python 3.8, Cuda 10.1, and PyTorch 1.8.1.The experiment is conducted on a Tesla V100-SXM2-32GB graphics card. Please ensure that all models within the experiment are configured identically, and the outcomes should reflect the mean result from a 5-fold cross-validation process. Four of these sections are used for learning one after the other, and the last one is used for checking the learning. We're gonna flip the images sideways and upside down to give it a good workout.The loss function is the Dice loss function. To stop the model from getting too cozy with the training data, we call it quits if it doesn't get better at its job in 10 tries.

B. *Datasets*

The dataset used in this experiment is the Lung Image Database Consortium and Image Database Resource Initiative (LIDC) [5]. The LIDC dataset is currently the world's largest public lung cancer dataset, which contains a total of 1,018 research cases, of which the CT lesion information is annotated by up to four doctors. This paper selects 1,186 lung nodule training samples according to the sample screening strategy of the LIDC dataset (CT layer thickness less than 2.5 mm, lung nodule diameter greater than 3 mm, and annotated by multiple doctors). The dataset example is shown in Figure 3.

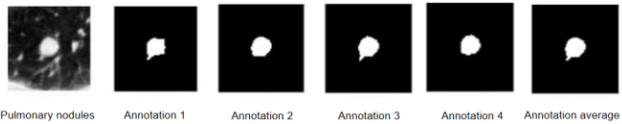

Figure 3 Dataset Example

### C. Experimental Results

Table 1 Experimental Results

| Model | Sample size | DSC | Sample selection strategy |
|---|---|---|---|
| SAVIC | 731 | 84.10 | Tube current: 40–582mA, slice thickness: 0.625-3.00mm |
| TUNET | 2638 | 84.23 | - |
| SONG | 751 | 85.11 | Slice thickness ≤2.5mm, lung nodule diameter ≥3mm |
| TYAGI | 2638 | 86.39 | - |
| DUNET | 2638 | 86.37 | - |
| UNETX | 751 | 87.21 | Slice thickness ≤2.5mm, lung nodule diameter ≥3mm |
| HUNET | 751 | 88.75 | Slice thickness ≤2.5mm, lung nodule diameter ≥3mm |
| ours | 751 | 89.10 | Slice thickness ≤2.5mm, lung nodule diameter ≥3mm |

As can be seen from Table 1, the SAVIC model obtained a DSC value of 84.10% with a small sample size (731 cases), and its sample selection strategy covered a wide range of tube current (40-582mA) and slice thickness (0.625-3.00mm). In contrast, although the TUNET model had the largest sample size (2638 cases), its DSC value was 84.23%, which was similar to the SAVIC model. It is worth noting that the TUNET model did not provide a specific sample selection strategy.

With a sample size of 751 cases, the DSC values of the five models, SONG, TYAGI, DUNET, UNETX, and HUNET, reached 85.11%, 86.39%, 86.37%, 87.21%, and 88.75%, respectively. These models all selected samples with slice thickness less than or equal to 2.5mm and lung nodule diameter greater than or equal to 3mm for training and testing, showing a relatively consistent sample selection strategy. Among them, the HUNET model performed best among these models.

Our model obtained a DSC value of 89.10% under the same sample selection strategy (i.e., slice thickness ≤ 2.5 mm, lung nodule diameter ≥ 3 mm), which is the highest score among all models, indicating that our model has the best segmentation performance under the specific sample selection strategy.

### D. Ablation experiment

This paper uses a simple CBR convolutional layer as the encoder-decoder basic block of the benchmark model and uses UNet and ResUNet as the comparison benchmark models. At the same time, this paper compares the performance of the BaseU, BaseRes, and BaseCR models. The design of the model is shown in Table 2.

Table 2. Channel residual and general basic blocks

| Model | SEN | DSC | PRE | mIOU |
|---|---|---|---|---|
| Unet | 85.13 | 86.77 | 82.03 | 83.33 |
| ResUnet | 82.85 | 87.65 | 82.21 | 83.65 |
| BaseU | 83.60 | 88.13 | 83.99 | 84.66 |
| BaseRes | 83.98 | 88.31 | 84.65 | 84.98 |
| BaseCR | 84.33 | 89.10 | 84.69 | 85.87 |

In this section, we will further verify the practicality of the proposed channel squeeze U-shaped structure. In this section, we build nine different models and compare and analyze them. At the same time, we pay special attention to some classic attention mechanism modules. These modules are integrated into the bottleneck layer of the network. In addition, we introduce the ResCBAM module, which adds the CBAM attention module to the ResNet basic block. The specific network structure is shown in Table 3.

Table 3 Verification of the channel extrusion U structure.

| Model | SEN | DSC | PRE | mIOU |
|---|---|---|---|---|
| Base4f | 85.13 | 81.77 | 82.03 | 84.73 |
| Baseceu | 82.85 | 82.65 | 82.21 | 85.43 |
| Base4f_ca | 83.60 | 83.13 | 82.99 | 85.76 |
| BaseResc | 83.88 | 83.31 | 83.65 | 85.76 |
| BaseCBAM | 83.93 | 84.10 | 83.77 | 85.77 |
| BaseSE | 84.01 | 84.23 | 83.32 | 86.03 |
| BaseSK | 84.13 | 84.32 | 84.01 | 86.11 |
| Base4f_res | 85.32 | 84.65 | 85.23 | 86.21 |
| ours | 86.04 | 84.78 | 84.15 | 86.37 |

According to the experimental results, we can observe the performance of different models in four key indicators: sensitivity (SEN), Dice similarity coefficient (DSC), precision (PRE), and mean intersection over union (mIOU). As can be seen from the table, the "Base4f" model performs best in SEN, reaching 85.13%, but is not optimal in the other three indicators. Although the "Baseceu" model has a lower SEN, it reaches 85.43% in mIOU, showing good overall performance. The "BaseSE" and "BaseSK" models show high scores in all evaluation indicators, especially "BaseSK", whose SEN, DSC, and PRE reach 84.13%, 84.32%, and 84.01% respectively, and mIOU reaches 86.11%, showing excellent comprehensive performance. The "Base4f_res" model reaches 85.32% and 86.21% in SEN and mIOU respectively, indicating that the model is not only highly sensitive but also has good generalization ability when dealing with complex scenes. Finally, our model achieved 86.04% and 86.37% on SEN and mIOU respectively. Although it was slightly lower than "Base4f_res" on PRE, overall, our model performed well in most indicators, especially in the key mIOU indicator, which showed that our model had good comprehensive performance and stability in image segmentation tasks.

## V. CONCLUSION

This paper introduces an advanced artificial intelligence-driven method, the "Channel Squeeze U-Structure," designed to significantly improve lung nodule detection and segmentation, thereby enhancing early-stage lung cancer diagnosis. By leveraging deep learning techniques, the model integrates key innovations such as shallow information processing, channel residual structures, and channel squeeze integration, enabling superior feature extraction and precise identification of small and complex nodules. The use of AI in this context not only boosts accuracy but also automates a labor-intensive task, supporting radiologists and mitigating the limitations of manual detection, especially in resource-constrained environments. While this method has demonstrated excellent performance on the LIDC dataset, future research should focus on incorporating multimodal imaging data and exploring state-of-the-art AI techniques, including Transformer-based architectures and attention mechanisms, to further optimize the model's ability to capture intricate lesion details. This AI-driven approach paves the way for the next generation of computer-aided diagnosis systems, offering transformative potential in the field of medical imaging and lung cancer detection.